# Effect of coupling with strain in multiferroics on phase diagrams and elastic anomalies.

E. V. Charnaya[*1], A. L. Pirozerskii[1], K. R. Gabbasova[1], A. S. Bugaev[2]

[1]*Physics Department, St. Petersburg State University, Petrodvorets, St. Petersburg, 198504 Russia*
[2]*Moscow Institute of Physics and Technology, Moscow, 141700 Russia*

The Landau theory was applied to treat the phase diagrams for a multiferroic with two second order phase transitions taking into account the coupling of the primary order parameters with strain. Two order parameters are coupled biquadratically which corresponds to the magnetoelectric materials. The coupling with strain is assumed to be linear in strain and quadratic in order parameters. Three ordered phases are discussed. Analytic relationships were obtained for the phase transition temperatures and for elastic modulus changes through the phase transitions. Strong influence of the coupling with strain on the phase diagrams was shown.

## 1. Introduction

Recently the renewed attention was focused on multiferroic materials (single-phase and composite ones) because of their promising applications as multifunctional devices (see [1,2] and references therein). The most interesting case is when two coupling order parameters are related to magnetic and electric orderings leading to the magnetoelectric effect. Experimental studies of multiferroic materials showed that their properties can be remarkably influenced by interaction with strains. Such an interaction becomes apparent, for instance, in elastic anomalies through the phase transitions [3-10] and in the impact of substrate-induced or epitaxial strain on the ferroic phase transitions and orientations of magnetic moments [11-13]. While many efforts were made to reveal the role of strain in the multiferroic materials, some effects of coupling of the magnetic and electric order parameters with strain were still not discussed properly.

In the present paper we will theoretically describe how the coupling with strain affects the phase diagrams and the magnitudes of the order parameters of a multiferroic using the phenomenological Landau approach. The Landau theory was repeatedly applied to treat the properties of multiferroic materials (see, for instance, [14,15]), however the treatment of the problem mentioned above was lacked. The precise expressions for the relevant elastic anomalies through the phase transitions will be also written for completeness, all the more they were not derived for some particular cases until now.

## 2. Phase diagrams

Let us consider a multiferroic which has two successive phase transitions associated with two primary order parameters $\eta$ and $\xi$. We restrict our discussion to the most interesting case of magnetic and electric order parameters, therefore the coupling between the order parameters takes the biquadratic form of $\frac{1}{2}\kappa\eta^2\xi^2$ ($\kappa$ is a phenomenological coupling constant). Additionally, we assume that both order parameters are coupled to strain $\varepsilon$ which plays the role of a secondary order parameter. While the phase diagram for the two coupled order parameters was considered repeatedly (see, for instance, [16,17]), its alterations caused by interaction with strain were never discussed. We will take into account the most important magnetoelastic coupling which is linear on strain (magnetostriction). Similarly, the electroelastic coupling is implied to be linear in strain and quadratic in spontaneous polarization (electrostriction). Then the Landau free energy expansion for the second order phase transitions can be written as

$$\Phi = \frac{1}{2}\alpha\eta^2 + \frac{1}{4}\beta\eta^4 + \frac{1}{2}\kappa\eta^2\xi^2 + \frac{1}{2}a\xi^2 + \frac{1}{4}b\xi^4 + \frac{1}{2}c\varepsilon^2 + \frac{1}{2}\theta_1\varepsilon\eta^2 + \frac{1}{2}\theta_2\varepsilon\xi^2, \qquad (1)$$

where $\alpha = \alpha_0(T - T_1)$, $a = a_0(T - T_2)$, $\alpha_0$, $a_0$, $\beta$, $b$, $c$, $\theta_1$, $\theta_2$ are phenomenological constants, $\alpha_0$, $a_0$, $\beta$, and $b$ are positive. We assume $T_2 > T_1$.

---
[*] Corresponding author: charnaya@live.com

The necessary conditions for $\Phi$ to have a local minimum at zero applied fields imply that the derivatives

$$\begin{cases} \dfrac{\partial \Phi}{\partial \eta} = \alpha\eta + \beta\eta^3 + \kappa\eta\xi^2 + \theta_1\varepsilon\eta \\[2mm] \dfrac{\partial \Phi}{\partial \xi} = a\xi + b\xi^3 + \kappa\eta^2\xi + \theta_2\varepsilon\xi \\[2mm] \dfrac{\partial \Phi}{\partial \varepsilon} = c\varepsilon + \dfrac{1}{2}\theta_1\eta^2 + \dfrac{1}{2}\theta_2\xi^2 \end{cases} \qquad (2)$$

are all equal zero (the equilibrium conditions). If we define for convenience

$$\tilde{\beta} = \beta - \frac{\theta_1^2}{2c}, \quad \tilde{b} = b - \frac{\theta_2^2}{2c}, \quad \tilde{\kappa} = \kappa - \frac{\theta_1\theta_2}{2c}, \qquad (3)$$

the equilibrium conditions can be written using (2) and (3) as

$$\begin{cases} \eta(\alpha + \tilde{\beta}\eta^2 + \tilde{\kappa}\xi^2) = 0 \\[2mm] \xi(a + \tilde{b}\xi^2 + \tilde{\kappa}\eta^2) = 0 \\[2mm] \varepsilon = -\dfrac{1}{2c}(\theta_1\eta^2 + \theta_2\xi^2) \end{cases}. \qquad (4)$$

The system (4) leads to the emergence of four different phases depending on phenomenological parameters and temperature: the paraphase $\eta = 0$, $\xi = 0$; two different ordered phases $\eta = 0$, $\xi \neq 0$ and $\eta \neq 0$, $\xi = 0$; multiferroic phase $\eta \neq 0$, $\xi \neq 0$. The conditions of stability of these four phases can be found using the Hesse matrix

$$A = \begin{pmatrix} \dfrac{\partial^2 \Phi}{\partial \eta^2} & \dfrac{\partial^2 \Phi}{\partial \xi \partial \eta} & \dfrac{\partial^2 \Phi}{\partial \varepsilon \partial \eta} \\[3mm] \dfrac{\partial^2 \Phi}{\partial \eta \partial \xi} & \dfrac{\partial^2 \Phi}{\partial \xi^2} & \dfrac{\partial^2 \Phi}{\partial \varepsilon \partial \xi} \\[3mm] \dfrac{\partial^2 \Phi}{\partial \eta \partial \varepsilon} & \dfrac{\partial^2 \Phi}{\partial \xi \partial \varepsilon} & \dfrac{\partial^2 \Phi}{\partial \varepsilon^2} \end{pmatrix} = \begin{pmatrix} \alpha + 3\beta\eta^2 + \kappa\xi^2 + \theta_1\varepsilon & 2\kappa\eta\xi & \theta_1\eta \\[2mm] 2\kappa\eta\xi & a + 3b\xi^2 + \kappa\eta^2 + \theta_2\varepsilon & \theta_2\xi \\[2mm] \theta_1\eta & \theta_2\xi & c \end{pmatrix}. \qquad (5)$$

A phase is stable when the Hesse matrix is positive definite. Then the free energy has a local minimum. To check whether the Hesse matrix is positive definite or not, one can use the Sylvester's criterion. According to this criterion, the eigenvalues of a symmetric matrix are all positive if and only if all leading principle minors are positive. Let us consider the ranges of existence of different phases individually.

(a) **Paraphase** $\left(\eta = \xi = \varepsilon = 0\right)$ **(phase 1).**

The Hesse matrix is diagonal with eigenvalues $\alpha$, $a$, and $c$. It is positive definite if $T > T_2$.

(b) **Phase** $\eta = 0, \xi \neq 0$ **(phase 2).**

From (4) the equilibrium values of the nonzero order parameters are given by $\xi^2 = -\dfrac{a}{\tilde{b}}$ and $\varepsilon = \dfrac{a\theta_2}{2\tilde{b}c}$. This phase can exist below $T_2$. The Hesse matrix

$$A = \begin{pmatrix} \alpha - \dfrac{\tilde{\kappa}a}{\tilde{b}} & 0 & 0 \\ 0 & 2b\xi^2 & \theta_2\xi \\ 0 & \theta_2\xi & c \end{pmatrix} \qquad (6)$$

is positive when

$$\begin{cases} c > 0 \\ det\begin{pmatrix} 2b\xi^2 & \theta_2\xi \\ \theta_2\xi & c \end{pmatrix} > 0 \\ det\, A > 0 \end{cases} . \qquad (7)$$

The second inequality in (7)

$$det\begin{pmatrix} 2b\xi^2 & \theta_2\xi \\ \theta_2\xi & c \end{pmatrix} = 2c\tilde{b}\xi^2 = -2ac > 0 \qquad (8)$$

is satisfied at $T < T_2$. The third inequality in (7) comes to

$$\alpha - \frac{\tilde{\kappa}a}{\tilde{b}} > 0 \qquad (9)$$

which is satisfied at $0 < T < T_2$ if $\tilde{\kappa} > \dfrac{\alpha_0\tilde{b}T_1}{a_0 T_2} \equiv \tilde{\kappa}_1$ and at $0 < \tilde{T}_{high} < T < T_2$ if $\tilde{\kappa} < \tilde{\kappa}_1$ where

$$\tilde{T}_{high} = T_2 - \frac{T_2 - T_1}{1 - \dfrac{a_0\tilde{\kappa}}{\alpha_0\tilde{b}}} . \qquad (10)$$

This means that the multiferroic phase may emergence only if $\tilde{\kappa} < \tilde{\kappa}_1$.

**(c) Phase $\eta \neq 0, \xi = 0$ (phase 3).**

From (4) we can find $\eta^2 = -\alpha/\tilde{\beta}$ and $\varepsilon = \dfrac{\alpha\theta_1}{2\tilde{\beta}c}$. The former imposes $T_1$ as the upper temperature limit for Phase 3. The Hesse matrix for Phase 3 is

$$A = \begin{pmatrix} 2\beta\eta^2 & 0 & \theta_1\eta \\ 0 & a - \dfrac{\tilde{\kappa}\alpha}{\tilde{\beta}} & 0 \\ \theta_1\eta & 0 & c \end{pmatrix} . \qquad (11)$$

One can show that the Hesse matrix (11) is positive definite within a temperature interval $0 < T < \tilde{T}_{low} < T_1$ only if $\tilde{\kappa} > \dfrac{a_0\tilde{\beta}}{\alpha_0}\dfrac{T_2}{T_1} \equiv \tilde{\kappa}_2$ where

$$\tilde{T}_{low} = T_1 - \frac{T_2 - T_1}{\dfrac{\tilde{\kappa}\alpha_0}{a_0\tilde{\beta}} - 1} . \qquad (12)$$

Phase 3 does not exist if $\tilde{\kappa} < \tilde{\kappa}_2$. The temperature intervals of existing Phases 2 and 3 must be not overlapped for the second order phase transitions. This requests the additional condition $\tilde{T}_{low} \leq \tilde{T}_{high}$ which is satisfied if

$$\tilde{\kappa}^2 \leq \tilde{b}\,\tilde{\beta} \,. \tag{13}$$

The inequality (13) imposes a restriction on the magnitude of the modified coupling constant $\tilde{\kappa}$ and can be denoted as the condition of the weak coupling.

### (d). Multiferroic phase $\eta \neq 0, \xi \neq 0$ (phase 4).

For this phase the Hesse matrix is

$$A = \begin{pmatrix} 2\beta\eta^2 & 2\kappa\xi\eta & \theta_1\eta \\ 2\kappa\xi\eta & 2b\xi^2 & \theta_2\xi \\ \theta_1\eta & \theta_2\xi & c \end{pmatrix}. \tag{14}$$

The Sylvester's criterion gives the following inequalities:

$$\begin{cases} 2\beta\eta^2 > 0 \\ det\begin{pmatrix} 2\beta\eta^2 & 2\kappa\xi\eta \\ 2\kappa\xi\eta & 2b\xi^2 \end{pmatrix} = 4(\beta b - \kappa^2)\eta^2\xi^2 > 0 \,. \\ det\, A > 0 \end{cases} \tag{15}$$

The first inequality in (15) is obviously satisfied. The second one leads to $\kappa^2 < \beta b$ which is the weak coupling condition for the case when there is no interaction with strain. It was assumed to be satisfied [17]. The third inequality in (15) can be written as

$$det\, A = \left[4c\left(b\beta - \kappa^2\right) - 2\theta_2\left(\beta\theta_2 - \kappa\theta_1\right) + 2\theta_1\left(\kappa\theta_2 - b\theta_1\right)\right]\xi^2\eta^2 = 4c\xi^2\eta^2\left(\tilde{b}\tilde{\beta} - \tilde{\kappa}^2\right), \tag{16}$$

which leads to the condition (13). Then the Hesse matrix for the multiferroic phase is positive definite under the conditions of weak coupling.

The equilibrium conditions for the multiferroic phase can be found from the system (4) which leads to

$$\begin{cases} \eta^2 = -\dfrac{\alpha\tilde{b} - \tilde{\kappa}a}{\tilde{\beta}\tilde{b} - \tilde{\kappa}^2} \\ \xi^2 = -\dfrac{a\tilde{\beta} - \tilde{\kappa}\alpha}{\tilde{\beta}\tilde{b} - \tilde{\kappa}^2} \\ \varepsilon = \dfrac{1}{2c}(\theta_1\dfrac{\alpha\tilde{b} - \tilde{\kappa}a}{\tilde{\beta}\tilde{b} - \tilde{\kappa}^2} + \theta_2\dfrac{a\tilde{\beta} - \tilde{\kappa}\alpha}{\tilde{\beta}\tilde{b} - \tilde{\kappa}^2}) \end{cases} \tag{17}$$

Taking into account the condition (13) and system (17) we can conclude that the multiferroic phase can exist in a temperature range defined by two inequalities

$$\begin{cases} \alpha - \dfrac{\tilde{\kappa}a}{\tilde{b}} < 0, \\ a - \dfrac{\tilde{\kappa}\alpha}{\tilde{\beta}} < 0. \end{cases} \tag{18}$$

Using the results for Phase 2 it is easy to see that the first inequality in (18) is satisfied only if $\tilde{\kappa} < \tilde{\kappa}_1$. In this case the temperature $\tilde{T}_{high}$ is positive and determines the upper boundary of the existence of Phase 4. Using the results for Phase 3 one can show that if $\tilde{\kappa} > \tilde{\kappa}_2$ then the multiferroic phase exists in the temperature range $\tilde{T}_{low} < T < \tilde{T}_{high}$. Below $\tilde{T}_{low}$ the multiferroic phase transforms into Phase 3. If $\tilde{\kappa} < \tilde{\kappa}_2$ the multiferroic phase exists in the temperature range $0 < T < \tilde{T}_{high}$ while Phase 3 does not emerge. As the condition of weak coupling should be satisfied, the case $\tilde{\kappa} > \tilde{\kappa}_2$ can happen only if $\tilde{\kappa}_2 < \tilde{\kappa}_m \equiv \sqrt{\tilde{\beta}\,\tilde{b}}$ .

It follows from the above consideration that the temperatures of the phase transitions and the magnitudes of the order parameters are strongly affected by interaction with elastic strain. Therefore, the phase diagrams in the presence of strain differ from those which normally are considered in publications when treating experimental data for multiferroic materials. It is of interest to compare the phase diagrams modified by strain with unmodified ones. Note that the unaffected values of the boundary temperatures can be found from (11) and (16) by replacing the parameters with the tilde by the relevant original parameters.

Fig.1 shows phase diagrams without and with coupling of the primary order parameters $\eta$ and $\xi$ with strain for particular sets of phenomenological parameters listed in the figure caption. The abscissa axis corresponds to the constant $\kappa$ which plays the role of a variable. Note that while $\kappa$ can vary between $-\sqrt{\beta b}$ and $\sqrt{\beta b}$ in the absence of coupling with strain, it varies between $\max(-\sqrt{b\beta},\ -\sqrt{\tilde{\beta}\tilde{b}} + \theta_1\theta_2/2c)$ and $\min(\sqrt{b\beta},\ \sqrt{\tilde{\beta}\tilde{b}} + \theta_1\theta_2/2c)$ in the presence of coupling with strain.

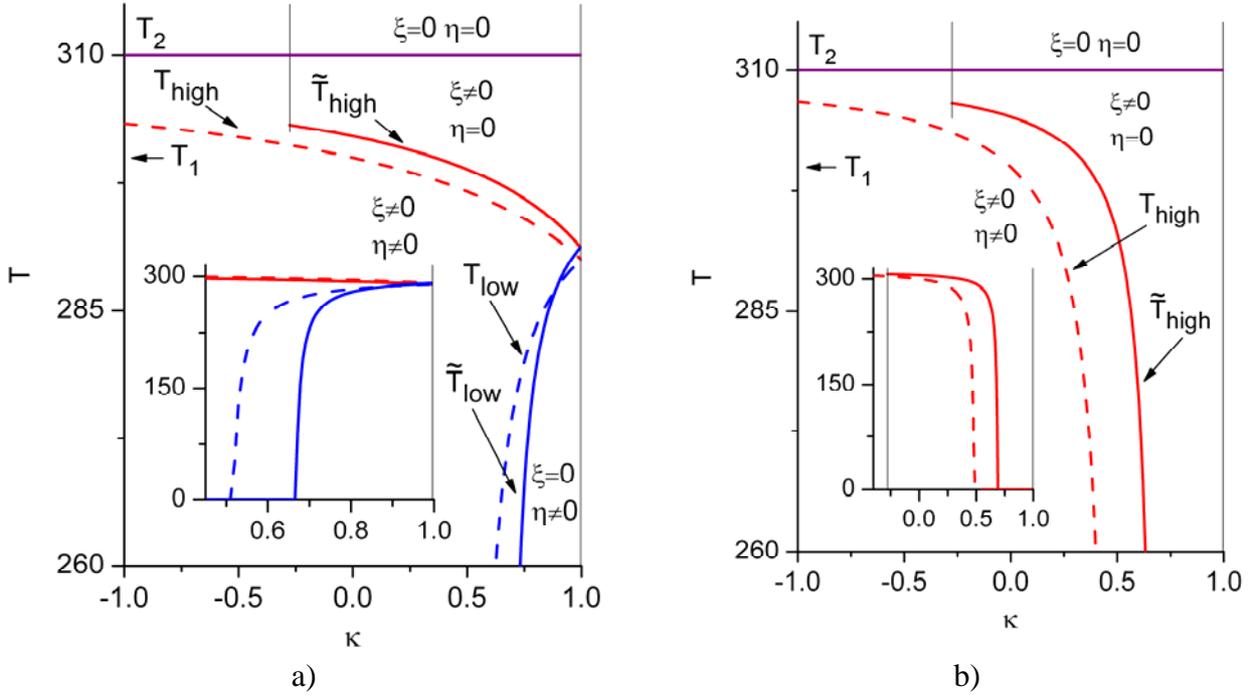

a)                                    b)

**Fig. 1.** Phase diagrams in the plane $\kappa - T$ with ($\theta_1 = 0.9$ , $\theta_2 = 0.8$; solid lines) and without ($\theta_1 = \theta_2 = 0$; dashed lines) coupling of the primary order parameters $\eta$ and $\xi$ with strain at higher temperatures. The phenomenological parameters of the Landau expansion are: $T_1$=300, $T_2$=310, $\beta$=1, $b$=1, $c$=1. Other parameters are: $\kappa_1 \cong 1.94$. $\kappa_2 \cong 0.52$, $\kappa_m = 1$, $\tilde{\beta} \cong 0.60$, $\tilde{b} \cong 0.68$, $\tilde{\kappa}_m \cong 0.64$. a) $\alpha_0$=0.01, $a_0$=0.005, $\tilde{\kappa}_1 \cong 1.32$, $\tilde{\kappa}_2 \cong 0.31$. b) $\alpha_0$=0.005, $a_0$=0.01, $\tilde{\kappa}_1 \cong 0.33$, $\tilde{\kappa}_2 \cong 1.23$. The horizontal solid line shows the upper boundary of Phase 2. The vertical thin lines mark the limits for admissible $\kappa$ under coupling with strain. The insets show the phase diagrams until zero temperature and for restricted ranges of $\kappa$ .

The left panel shows the phase diagrams for the case $\kappa_2 < \kappa_m$ and $\tilde{\kappa}_2 < \tilde{\kappa}_m$ . In this case Phase 3 can emerge in the intervals $\kappa_2 < \kappa < \kappa_m$ and $\tilde{\kappa}_2 < \tilde{\kappa} < \tilde{\kappa}_m$ without and with coupling with strain, respectively. Not that the magnitudes of $\kappa_1$ and $\tilde{\kappa}_1$ satisfy the inequalities $\kappa_1 > \kappa_m$ and $\tilde{\kappa}_1 > \tilde{\kappa}_m$ since $\kappa_1 \kappa_2 = \kappa_m^2$ and $\tilde{\kappa}_1 \tilde{\kappa}_2 = \tilde{\kappa}_m^2$. Therefore, the multiferroic phase can exist at any admissible $\kappa$ and $\tilde{\kappa}$.

The right panel corresponds to the case $\kappa_2 > \kappa_m$ and $\tilde{\kappa}_2 > \tilde{\kappa}_m$. Phase 3 does not emerge at any admissible $\kappa$ and $\tilde{\kappa}$. The magnitudes of $\kappa_1$ and $\tilde{\kappa}_1$ satisfy the inequalities $\kappa_1 < \kappa_m$ and $\tilde{\kappa}_1 < \tilde{\kappa}_m$. Then the multiferroic phase does not exist at $\kappa_1 < \kappa < \kappa_m$ and $\tilde{\kappa}_1 < \tilde{\kappa} < \tilde{\kappa}_m$ without and with coupling with strain, respectively.

Phase diagrams are very sensitive to changes in the phenomenological coefficients $\theta_1$ and $\theta_2$ which characterize the coupling of the primary order parameters with strain. Fig.2 shows variations of the boundary temperatures $\tilde{T}_{high}$ and $\tilde{T}_{low}$ with $\theta_1$ at different $\theta_2$ and $\kappa$.

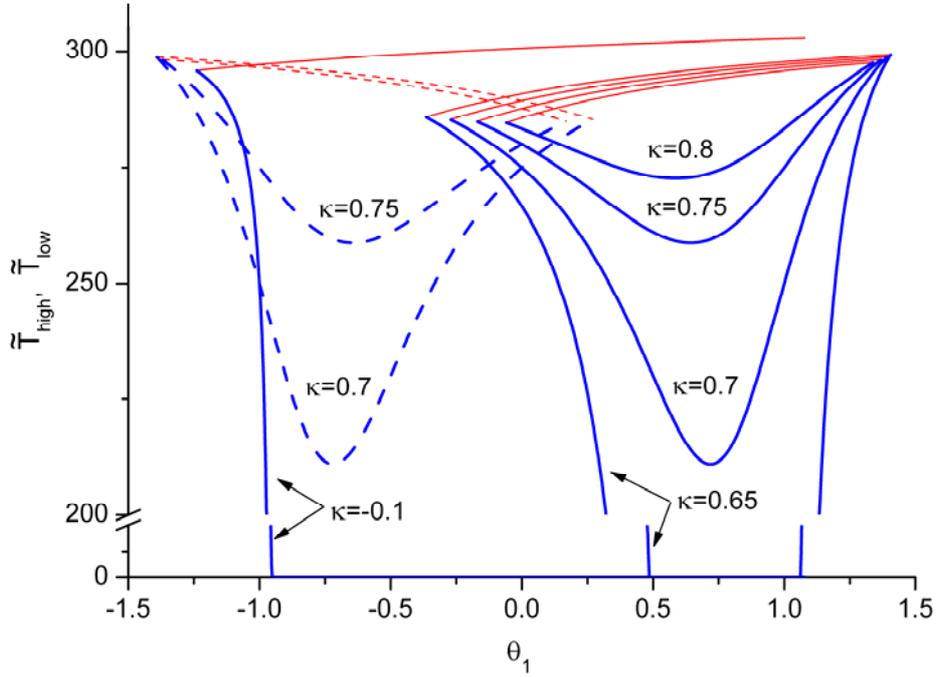

**Fig. 2.** Dependence of the temperatures $\tilde{T}_{high}$ (thin lines, red online) and $\tilde{T}_{low}$ (bold lines, blue online) on $\theta_1$ at $\theta_2$ =0.8 (solid lines) and $\theta_2$ =-0.8 (dashed lines) and different $\kappa$ indicated in the panel. Other phenomenological parameters of the Landau expansion are: $T_1$=300, $T_2$=310, $\alpha_0$=0.01, $a_0$=0.005, $\beta$=1, $b$=1, $c$=1.

One can see that different sequences of phases can be expected depending on the strength of coupling with strain. For instance, the temperature interval of the emergence of multiferroic phase at $\theta_2$ =0.8 and $\kappa$ =0.8 is broader near $\theta_1$ =0.65 while the multiferroic phase disappears for $\theta_1$ <-0.05.

Fig.3 shows the variations with temperature of the magnitudes of the primary order parameters $\eta$ and $\xi$ for several different values of $\theta_1$ and $\theta_2$. One can see that the upper and lower magnitudes of the primary order parameters are affected but slightly and the most noticeable changes in their variations with temperature arise due to alterations in the temperatures of the phase transitions.

## 3. Elastic modulus.

In addition to the influence on the phase diagrams, the coupling of the magnetic and electric order parameters with strain leads to the emergence of elastic anomalies at modified phase transition temperatures. The alterations in the elastic modulus can be monitored by ultrasound techniques as in [3-10]. Elastic anomalies through the phase transitions in multiferroics were discussed previously [5,7] within the framework of the Landau theory. However, expressions for the elastic modulus were obtained on the basis of a truncated Landau expansion which was written for only one order parameter coupled with strain. Here we will obtain the expressions for the elastic modulus changes through the phase transitions in multiferroics within the complete expansion (1).

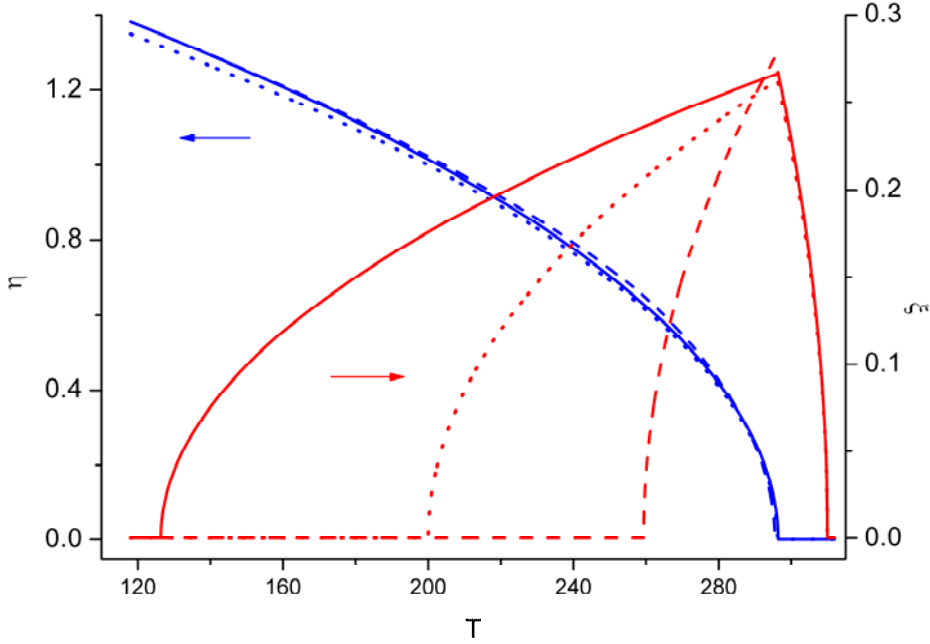

**Fig. 3.** Temperature dependences of the magnitudes of the primary order parameters $\eta$ (blue online) and $\xi$ (red online). Dotted lines: without coupling with strain ($\theta_1 = \theta_2 = 0$). Solid lines: $\theta_1 = \theta_2 = 0.3$. Dashed lines: $\theta_1 = -\theta_2 = 0.3$. $\kappa = 0.55$ and other phenomenological parameters of the Landau expansion are the same as in Fig.2.

Let us find the relationships for the elastic modulus in Phases 2-4. Note that in the paraphrase where the spontaneous order parameters are all equal zero, the elastic modulus is $c$. The elastic modulus in other phases can be found as

$$c_{eff} \equiv \left( \frac{\partial \varepsilon}{\partial \sigma} \right)^{-1},\qquad(19)$$

where $\sigma$ is a stress associated with $\varepsilon$. The derivative (19) must be calculated in the zero stress limit. To find the derivative (19) we have to write the thermodynamic potential (1) in the presence of stress adding to the right side of (1) the term $-\varepsilon\sigma$. In this case the derivative $\frac{\partial \Phi}{\partial \varepsilon}$ becomes

$\frac{\partial \Phi}{\partial \varepsilon} = c\varepsilon + \frac{1}{2}\theta_1\eta^2 + \frac{1}{2}\theta_2\xi^2 - \sigma$ while other derivatives in (2) are not changed. Differentiating $\frac{\partial \Phi}{\partial \eta}$, $\frac{\partial \Phi}{\partial \xi}$, and $\frac{\partial \Phi}{\partial \varepsilon}$ with respect to $\sigma$ we can easily find the following matrix equation

$$A\begin{pmatrix} \frac{\partial \eta}{\partial \sigma} \\ \frac{\partial \xi}{\partial \sigma} \\ \frac{\partial \varepsilon}{\partial \sigma} \end{pmatrix} = \begin{pmatrix} 0 \\ 0 \\ 1 \end{pmatrix},\qquad(20)$$

where A is the Hesse matrix (5). From (20) using the Cramer's rule we immediately find that in Phase 2

$$c_{eff} = c\frac{\tilde{b}}{b},\qquad(21)$$

in Phase 3

$$c_{eff} = c\frac{\tilde{\beta}}{\beta},\qquad(22)$$

and in the multiferroic phase

$$c_{eff} = c\frac{\tilde{b}\tilde{\beta} - \tilde{\kappa}^2}{b\beta - \kappa^2}.\qquad(23)$$

The relationships (21)-(23) predict step-like changes in the elastic modulus at the transitions between different phases. An example of the temperature dependence of the relative changes in the elastic modulus $\Delta c_{eff}/c = (c_{eff} - c)/c$ through the phase transitions for the same phenomenological coefficients as in Fig.3 is shown in Fig.4. Note that the elastic modulus decreases with decreasing temperature through the phase transitions into Phase 2 and multiferroic phase, however it increases at the transition into Phase 3. In real materials, the step-like changes must be smeared which leads to gradual decrease or increase in the elastic modulus through the phase transitions.

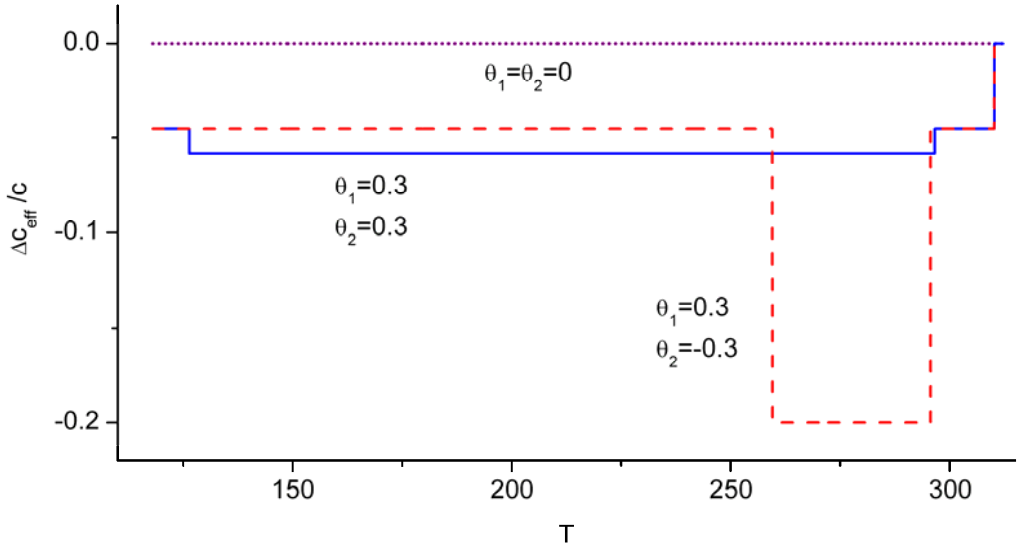

**Fig. 4.** Temperature dependences of the elastic modulus. The coupling constants of the primary order parameters with strain are $\theta_1 = \theta_2 = 0$ (dotted line, purple online), $\theta_1 = \theta_2 = 0.3$ (solid line, blue online), $\theta_1 = -\theta_2 = 0.3$ (dashed line, red online). Other phenomenological parameters of the Landau expansion are the same as in Fig.3.

In general, the coupling of the primary order parameters with strain can also include a biquadratic terms like $\lambda_1\varepsilon^2\eta^2$ and $\lambda_2\varepsilon^2\xi^2$ ($\lambda_1$ and $\lambda_2$ are phenomenological constants). Such additional coupling leads to monotonic temperature dependences of the elastic modulus in the ordered phases which can explain experimental results on ultrasound velocity [3-5,8]. However, if the biquadratic terms are added to the relationship (1), the analytic solutions for the elastic modulus cannot be found. The addition of the biquadratic coupling affects also the phase diagrams, but this influence is not significant.

In conclusion, phase diagrams are discussed analytically within the framework of the Landau approach for a magnetoelectric with two successive phase transitions associated with two primary order parameters coupled with strain. Strong influence of the coupling with strain on the emergence of particular phases (including the multiferroic one) and on alterations in the temperatures of the phase transitions between the paraphase and three ordered phases was found. In addition, accurate analytic expressions for the elastic modulus were derived.

**References.**

1. K. F. Wang, J. M. Liu, Z. F. Ren, *Adv. Phys.* 58, 321 (**2009**).
2. J. Ma, J. Hu, Z. Li, and C.-W. Nan, *Adv. Mater. 23*, 1062, (**2011).**
3. V. Felea, P. Lemmens, S. Yasin, S. Zherlitsyn, K. Y. Choi, C. T. Lin, and C. Payen, J. Phys.: Condens. Matter 23, 216001 (2011).
4. M. Poirier, J. C. Lemyre, P.-O. Lahaie, L. Pinsard-Gaudart, and A. Revcolevschi, Phys. Rev. B 83, 054418 (2011).
5. M. Poirier, F. Laliberté, L. Pinsard-Gaudart, and A. Revcolevschi, Phys. Rev. B **76**, 174426 (2007).
6. R. Villarreal, G. Quirion, M. L. Plumer, M. Poirier, T. Usui, and T. Kimura, Phys. Rev. Lett. 109, 167206 (2012).
7. G. Quirion, M. J. Tagore, M. L. Plumer, and O. A. Petrenko, Phys. Rev. B 77, 094111 (2008).
8. G. Quirion, M. L. Plumer, O. A. Petrenko, G. Balakrishnan, and C. Proust Phys. Rev. B 80, 064420 (2009).
9. E. Smirnova, A. Sotnikov, S. Ktitorov, N. Zaitseva, H. Schmidt, and M. Weihnacht, Eur. Phys. J. B 83, 39 (2011).
10. M. Kinka, V. Samulionis, J. Banys, A. Kalvane, and K. Bormanis, Ferroelectrics, 440, 93 (2012).
11. Y. S. Hou, J. H. Yang, X. G. Gong, and H. J. Xiang, Phys. Rev. B 88, 060406 (2013).
12. J. Grafe, M. Welke, F. Bern, M. Ziese, and R. Denecke, J. Magn. Magn. Mater. 339, 84 (2013).
13. Z. V. Gareeva, A. F. Popkov, S. V. Soloviov, and A. K. Zvezdin, Phys. Rev. B 87 214413 (2013).
14. E. K. H. Salje and M. A. Carpenter, J. Phys.: Condens. Matter 23, 462202 (2011).
15. P. Tolédano, N. Leo, D. D. Khalyavin, L. C. Chapon, T. Hoffmann, D. Meier, and M. Fiebig, Phys. Rev. Lett. 106, 257601 (2011).
16. V. E. Yurkevich, B. N. Rolov, and H. E. Stanley, Ferroelectrics, 16, 61 (1977).
17. Y. A. Izumov and V. N. Syromyatnikov, Phase transitions and crystal symmetry, Dordrecht, Netherlands; Boston: Kluwer Academic Publishers, 1990.